\documentclass[twocolumn,prc,superscriptaddress,showpacs,amssymb,amsmath,amsfonts,aps,floatfix]{revtex4}
\usepackage{graphicx}
\usepackage{amsmath}
\usepackage{color}
\usepackage{bm}	
\usepackage{xcolor}
\usepackage{hyperref}
\usepackage{times}
\usepackage{subfigure}

\newcommand{\ra}[1]{\rightarrow{#1}}
\def \lq{\textquotedblleft}
\def \rq{\textquotedblright~}

\hypersetup{
	colorlinks,
	citecolor=blue,
	linkcolor=blue
}

\hypersetup{
	urlbordercolor=white,
}
\begin{document}
\title{Transverse Polarization of $\Sigma^{+}(1189)$ in Photoproduction on a Hydrogen Target in CLAS}

\newcommand*{\ANL}{Argonne National Laboratory, Argonne, Illinois 60439}
\newcommand*{\ANLindex}{1}
\affiliation{\ANL}
\newcommand*{\ASU}{Arizona State University, Tempe, Arizona 85287-1504}
\newcommand*{\ASUindex}{2}
\affiliation{\ASU}
\newcommand*{\CSUDH}{California State University, Dominguez Hills, Carson, CA 90747}
\newcommand*{\CSUDHindex}{3}
\affiliation{\CSUDH}
\newcommand*{\CMU}{Carnegie Mellon University, Pittsburgh, Pennsylvania 15213}
\newcommand*{\CMUindex}{4}
\affiliation{\CMU}
\newcommand*{\CUA}{Catholic University of America, Washington, D.C. 20064}
\newcommand*{\CUAindex}{5}
\affiliation{\CUA}
\newcommand*{\SACLAY}{CEA, Centre de Saclay, Irfu/Service de Physique Nucl\'eaire, 91191 Gif-sur-Yvette, France}
\newcommand*{\SACLAYindex}{6}
\affiliation{\SACLAY}
\newcommand*{\CNU}{Christopher Newport University, Newport News, Virginia 23606}
\newcommand*{\CNUindex}{7}
\affiliation{\CNU}
\newcommand*{\UCONN}{University of Connecticut, Storrs, Connecticut 06269}
\newcommand*{\UCONNindex}{8}
\affiliation{\UCONN}
\newcommand*{\EDINBURGH}{Edinburgh University, Edinburgh EH9 3JZ, United Kingdom}
\newcommand*{\EDINBURGHindex}{9}
\affiliation{\EDINBURGH}
\newcommand*{\FU}{Fairfield University, Fairfield CT 06824}
\newcommand*{\FUindex}{10}
\affiliation{\FU}
\newcommand*{\FIU}{Florida International University, Miami, Florida 33199}
\newcommand*{\FIUindex}{11}
\affiliation{\FIU}
\newcommand*{\FSU}{Florida State University, Tallahassee, Florida 32306}
\newcommand*{\FSUindex}{12}
\affiliation{\FSU}
\newcommand*{\Genova}{Universit$\grave{a}$ di Genova, 16146 Genova, Italy}
\newcommand*{\Genovaindex}{13}
\affiliation{\Genova}
\newcommand*{\GWUI}{The George Washington University, Washington, DC 20052}
\newcommand*{\GWUIindex}{14}
\affiliation{\GWUI}
\newcommand*{\ISU}{Idaho State University, Pocatello, Idaho 83209}
\newcommand*{\ISUindex}{15}
\affiliation{\ISU}
\newcommand*{\INFNFE}{INFN, Sezione di Ferrara, 44100 Ferrara, Italy}
\newcommand*{\INFNFEindex}{16}
\affiliation{\INFNFE}
\newcommand*{\INFNFR}{INFN, Laboratori Nazionali di Frascati, 00044 Frascati, Italy}
\newcommand*{\INFNFRindex}{17}
\affiliation{\INFNFR}
\newcommand*{\INFNGE}{INFN, Sezione di Genova, 16146 Genova, Italy}
\newcommand*{\INFNGEindex}{18}
\affiliation{\INFNGE}
\newcommand*{\INFNRO}{INFN, Sezione di Roma Tor Vergata, 00133 Rome, Italy}
\newcommand*{\INFNROindex}{19}
\affiliation{\INFNRO}
\newcommand*{\ORSAY}{Institut de Physique Nucl\'eaire ORSAY, Orsay, France}
\newcommand*{\ORSAYindex}{20}
\affiliation{\ORSAY}
\newcommand*{\ITEP}{Institute of Theoretical and Experimental Physics, Moscow, 117259, Russia}
\newcommand*{\ITEPindex}{21}
\affiliation{\ITEP}
\newcommand*{\JMU}{James Madison University, Harrisonburg, Virginia 22807}
\newcommand*{\JMUindex}{22}
\affiliation{\JMU}
\newcommand*{\KNU}{Kyungpook National University, Daegu 702-701, Republic of Korea}
\newcommand*{\KNUindex}{23}
\affiliation{\KNU}
\newcommand*{\LPSC}{LPSC, Universite Joseph Fourier, CNRS/IN2P3, INPG, Grenoble, France
}
\newcommand*{\LPSCindex}{24}
\affiliation{\LPSC}
\newcommand*{\UNH}{University of New Hampshire, Durham, New Hampshire 03824-3568}
\newcommand*{\UNHindex}{25}
\affiliation{\UNH}
\newcommand*{\NSU}{Norfolk State University, Norfolk, Virginia 23504}
\newcommand*{\NSUindex}{26}
\affiliation{\NSU}
\newcommand*{\OHIOU}{Ohio University, Athens, Ohio  45701}
\newcommand*{\OHIOUindex}{27}
\affiliation{\OHIOU}
\newcommand*{\ODU}{Old Dominion University, Norfolk, Virginia 23529}
\newcommand*{\ODUindex}{28}
\affiliation{\ODU}
\newcommand*{\RPI}{Rensselaer Polytechnic Institute, Troy, New York 12180-3590}
\newcommand*{\RPIindex}{29}
\affiliation{\RPI}
\newcommand*{\MSU}{Skobeltsyn Nuclear Physics Institute, 119899 Moscow, Russia}
\newcommand*{\MSUindex}{30}
\affiliation{\MSU}
\newcommand*{\SCAROLINA}{University of South Carolina, Columbia, South Carolina 29208}
\newcommand*{\SCAROLINAindex}{31}
\affiliation{\SCAROLINA}
\newcommand*{\JLAB}{Thomas Jefferson National Accelerator Facility, Newport News, Virginia 23606}
\newcommand*{\JLABindex}{32}
\affiliation{\JLAB}
\newcommand*{\UTFSM}{Universidad T\'{e}cnica Federico Santa Mar\'{i}a, Casilla 110-V Valpara\'{i}so, Chile}
\newcommand*{\UTFSMindex}{33}
\affiliation{\UTFSM}
\newcommand*{\GLASGOW}{University of Glasgow, Glasgow G12 8QQ, United Kingdom}
\newcommand*{\GLASGOWindex}{34}
\affiliation{\GLASGOW}
\newcommand*{\VIRGINIA}{University of Virginia, Charlottesville, Virginia 22901}
\newcommand*{\VIRGINIAindex}{35}
\affiliation{\VIRGINIA}
\newcommand*{\WM}{College of William and Mary, Williamsburg, Virginia 23187-8795}
\newcommand*{\WMindex}{36}
\affiliation{\WM}
\newcommand*{\YEREVAN}{Yerevan Physics Institute, 375036 Yerevan, Armenia}
\newcommand*{\YEREVANindex}{37}
\affiliation{\YEREVAN}

\newcommand*{\NOWINFNGE}{INFN, Sezione di Genova, 16146 Genova, Italy}
\newcommand*{\NOWCNU}{Christopher Newport University, Newport News, Virginia 23606}
\newcommand*{\NOWLANL}{Los Alamos National Laboratory, Los Alamos, NM 87544 USA}
\newcommand*{\NOWMSU}{Skobeltsyn Nuclear Physics Institute, 119899 Moscow, Russia}
\newcommand*{\NOWORSAY}{Institut de Physique Nucl\'eaire ORSAY, Orsay, France}
\newcommand*{\NOWROMAII}{Universita' di Roma Tor Vergata, 00133 Rome Italy}

\author{C.S. Nepali\footnote{Contact author: \href{}{cnepali@jlab.org}}}
\affiliation{\ODU}
\author{M. Amaryan}
\affiliation{\ODU}
\author{K.P. ~Adhikari} 
\affiliation{\ODU}
\author{M.~Aghasyan} 
\affiliation{\INFNFR}
\author{S. ~Anefalos~Pereira} 
\affiliation{\INFNFR}
\author{H.~Baghdasaryan} 
\affiliation{\VIRGINIA}
\affiliation{\ODU}
\author{J.~Ball} 
\affiliation{\SACLAY}
\author{M.~Battaglieri} 
\affiliation{\INFNGE}
\author{V.~Batourine} 
\affiliation{\JLAB}
\affiliation{\KNU}
\author{I.~Bedlinskiy} 
\affiliation{\ITEP}
% ...

\collaboration{The CLAS Collaboration}
\noaffiliation

 \date{\today}
%
%
%
%
%
%
%

%
%
\begin{abstract}
	Experimental results on the $\Sigma^+(1189)$ hyperon transverse polarization in photoproduction on a hydrogen target using
	the CLAS detector at Jefferson laboratory are presented. The $\Sigma^+(1189)$ was reconstructed in the exclusive reaction $\gamma+p\rightarrow K^{0}_{S} + \Sigma^+(1189)$ via 
	the $\Sigma^{+} \rightarrow p \pi^{0}$ decay mode. The $K^{0}_S$ was reconstructed in the invariant mass of two oppositely charged
	pions with the $\pi^0$ identified in the missing mass of the detected $p\pi^+\pi^-$ final state. 
	Experimental data were collected in the photon energy range $E_{\gamma}$ = 1.0-3.5 GeV ($\sqrt{s}$ range 1.66-2.73 GeV). We observe a large negative polarization 
	of up to 95$\%$. As the mechanism of  transverse  polarization of hyperons produced in unpolarized photoproduction experiments is still 
	not well understood, these results will help to distinguish between different theoretical models on hyperon production and provide 
	valuable information for the searches of missing baryon resonances.
\end{abstract}
	\maketitle
\section{Introduction}
\label{intro}
	The constituent quark model is very successful in describing the observed baryon states.  
	However, there are a number of predicted 
	baryon states that have never been observed, {\it i.e.}, the \lq missing resonance\rq problem~\cite{Isgur:1978xj}. Predictions suggest 
	that some of these states decay primarily to hyperon-kaon ($YK$) final states~\cite{Capstick:1992uc}. This has initiated intense experimental activity 
	in photoproduction 
	of these channels at facilities such as SAPHIR, GRAAL, and JLab-CLAS. The main results were obtained in the reactions 
	$\gamma p\to \Lambda K^+$, $\gamma p\to \Sigma^0 K^+$, and $\gamma p\to \Sigma^+K_S^0$%
	~\cite{Schumacher:2008xw,McNabb:2003nf,Lleres:2007tx,Glander:2003jw,Bradford:2005pt,Bradford:2006ba,McCracken:2009ra,Dey:2010hh}. 
	Recently, several new resonances have been shown to exist~\cite{Anisovich:2011ye,Anisovich:2012ct} at around 2 GeV based on a multichannel partial-wave analysis of 
	existing data on pion- and photon-induced inelastic reactions. 
	In those reactions, hyperons were seen to be polarized normal to the production plane (a plane made by the momentum vector of the beam and the momentum
	vector of the hyperon, {\it i.e.}, along $\hat{n}_{z} = \hat{p}_{\text{beam}} \times \hat{p}_{\text{hyperon}}/|\hat{p}_{\text{beam}} \times \hat{p}_{\text{hyperon}}|$) 
	although neither beam nor target were polarized. 
	The study of hyperon polarization gives an important insight into the mechanism of $s\bar{s}$ pair creation, including the $s$-quark polarization with 
	subsequent polarization transfer to the produced hyperons~\cite{Carman:2002se,Carman:2009fi}. Because the hyperon polarization is a result of the interference 
	between the spin dependent
	and spin independent parts of the scattering amplitude~\cite{Sakurai:1964}, its experimental study provides access to various amplitudes contributing to 
	the production of hyperons~\cite{Heller:1990gc}.

	CLAS has measured $\Lambda$ and $\Sigma^{0}$ polarization with the highest statistical precision so far up to $\sqrt{s} \approx 2.84$ 
	GeV~\cite{McNabb:2003nf,McCracken:2009ra,Dey:2010hh}. Based on a simple non-relativistic quark model
	the $ud$ quark-pair wave function in the $\Lambda$ is
	anti-symmetric in both flavour and spin, and as a result, this quark pair does not carry a spin. Therefore, the $\Lambda$ polarization is given by the 
	strange quark. However, the $ud$ quark pair in the $\Sigma^{0}$ is in a spin 1 state pointing in the direction of the $\Sigma^{0}$ spin. Then the spin $\frac{1}{2}$
	of the $\Sigma^{0}$ is due to the opposite direction of the strange quark spin. 
	The $s$ quark is either produced polarized or else acquires it during recombination with the incident baryon fragments.
	Hence the polarization of the $\Lambda$ and the $\Sigma^{0}$ should be 
	similar in magnitude but opposite in direction. However, recent CLAS results~\cite{McNabb:2003nf,Dey:2010hh} show that while this symmetry, 
	$P_{\Lambda} \approx -P_{\Sigma^{0}}$, holds for backward production angles of the hyperon in the center-of-mass (CM), it is broken for mid and forward 
	hyperon production
	angles in the CM. 
	For the case of the $\Sigma^+$, we should expect that $P_{\Sigma^+} \approx P_{\Sigma^0}$ based on isospin symmetry when comparing the reactions 
	$\gamma p \to \Sigma^+ K_S^0$ and $\gamma p \to \Sigma^0 K^+$.

	Polarization of the $\Sigma^{+} (1189)$ in photoproduction on a proton target has been measured by SAPHIR~\cite{Lawall:2005np} but statistics are low and the $\Sigma^{+}$
	polarization was measured in a limited kinematic range. The measurement of polarization of all hyperons with higher statistics compared to the present world data is needed 
	to better understand the mechanism of $s\bar{s}$ quark pair creation and subsequent $s$ quark polarization.

	Below we present experimental results on the transverse polarization of the $\Sigma^{+}$ hyperon from the reaction $\gamma p \to \Sigma^{+} K^{0}_{S}$ obtained 
	with an unpolarized tagged photon beam and an unpolarized hydrogen target with  CLAS in the photon beam energy range 1.0-3.5 GeV (which corresponds to 
	$\sqrt{s} \approx$ 1.66-2.73 GeV) with higher statistics compared to the available world data so far.
\section{Experiment}
\label{method}
	The experiment was carried out using the CLAS detector~\cite{Mecking:2003zu} and the Hall-B photon tagging facility~\cite{Sober:2000we}.
	The photon beam is produced by bremsstrahlung  of unpolarized electrons in a thin gold foil radiator of thickness $10^{-4}$ radiation lengths. 
	The photon energy tagging range is from 20\% to 95\% of the incident electron energy~\cite{Sober:2000we}. 
	The target cell was 40~cm long, placed 10~cm upstream of the nominal CLAS center. Additional details of the experimental setup and the
	CLAS detector can be found in~\cite{Mecking:2003zu}. 

	We are using events with $\Sigma^{+}$(1189) produced via the following reaction 
	\begin{eqnarray}
		\gamma + p \ra K^{0} + \Sigma^{+}, 
	\end{eqnarray}
	where the $K^{0}$ is a mixture of $K^{0}_{S}$ and $K^{0}_L$~\cite{GellMann:1955jx,Pais:1955sm} 
	which are CP eigenstates:
	\begin{eqnarray}  
		|K^{0}\textgreater = \frac{1}{\sqrt{2}} (|K^{0}_{S}\textgreater ~+~ |K^{0}_{L}\textgreater).
	\end{eqnarray}
	Then, the $K^{0}_{S}$, being a short lived meson, decays quickly to $\pi^{+}$ and $\pi^{-}$ with a branching ratio of 69\%~\cite{pdg} via 
	the CP conserving weak decay, while the $K^{0}_{L}$, being a long lived meson, decays essentially beyond the CLAS detector, which makes it undetectable. 
	The $\Sigma^{+}$ decays to a proton and $\pi^{0}$ with a branching ratio of 51\%~\cite{pdg} via weak decay. So, the detected final state 
	particles are proton, $\pi^{+}$, and $\pi^{-}$, while the $\pi^{0}$ is reconstructed from the missing mass of the proton and $K^{0}_{S}$. 
	The $K^{0}_{S}$ is reconstructed from the invariant mass of $\pi^+\pi^-$:
	\begin{eqnarray}
		\gamma + p \rightarrow K^{0}_{S} + \Sigma^{+}    
			   \rightarrow \pi^{+} + \pi^{-} + p + \pi^{0}.
		\label{equ:channel}
	\end{eqnarray}
	The $\Sigma^+$ is reconstructed in the missing mass of $K^{0}_S$ by requiring the missing mass of the proton and $K^{0}_{S}$ to be $\pi^0$.
\section{Event Selection}
\label{event}
	Charged particles were identified by the time-of-flight method and their momenta. Their momenta were obtained from tracking in the drift chambers.  
	Events were selected if they contained one and only one $p$, $\pi^{+}$ and $\pi^{-}$. 
	The photon, whose arrival time at the interaction vertex as measured by the photon tagging system was closest to the event start time
	measured in CLAS, was selected as the photon that initiated the reaction. Selected events should have only one photon detected in the photon
	tagging system within $\pm1$ ns of the event in the CLAS because the time interval between electron beam buckets~\cite{Sober:2000we} is 2 ns.
	A correction was applied to the photon energy that accounts for mechanical distortion of the photon tagging plane, and  energy loss and momentum corrections 
	were applied to all detected charged particles by using the CLAS energy loss and momentum correction packages~\cite{CLAS-NOTE:2005-002}.

\begin{figure}[hbt]
		\begin{center}
			\resizebox{6.5cm}{!}{\includegraphics*{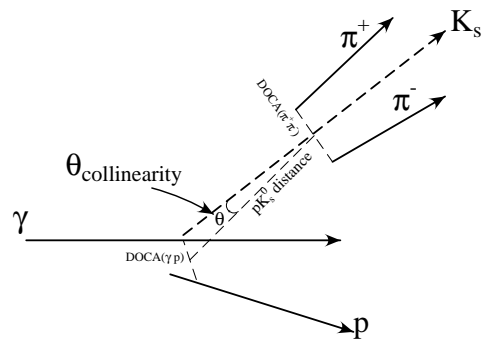}}
				\caption{Different distances of closest approach and the collinearity angle, $\theta_{\text{collinearity}}$. 
				DOCA($\pi^{+}\pi^{-}$) is the distance of closest approach between the two charged pions. DOCA($\gamma p$) is the distance
				of closest approach between the photon and the proton. \lq$pK^{0}_{s}$ distance\rq is the distance between the center of DOCA($\gamma p$) and 
				the center of DOCA($\pi^{+}\pi^{-}$).}
			\label{fig:dist}
		\end{center}
	\end{figure}

\begin{figure}[hbt]
		\begin{center}
			\resizebox{7.5cm}{!}{\includegraphics*{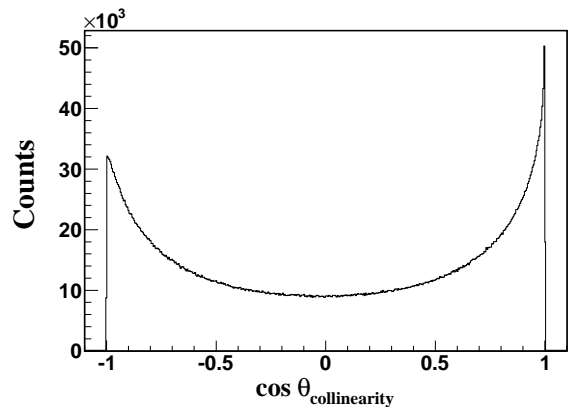}}
				\caption{Distribution of cosine of the collinearity angle, $\cos\theta_{\text{collinearity}}$, with 
				the $\pi^{0}$ and $K^{0}_{S}$ selected. This distribution results from the DOCA resolutions (Fig.~\ref{fig:dist}), the $K^{0}_{S}$ decay
				distance distribution, and the non-resonant $\pi\pi$ continuum.}
			\label{fig:coll}
		\end{center}
	\end{figure}

\begin{figure*}[hbt]
		\begin{center}
			\subfigure[] 
			{
				\resizebox{7.5cm}{!}{\includegraphics*{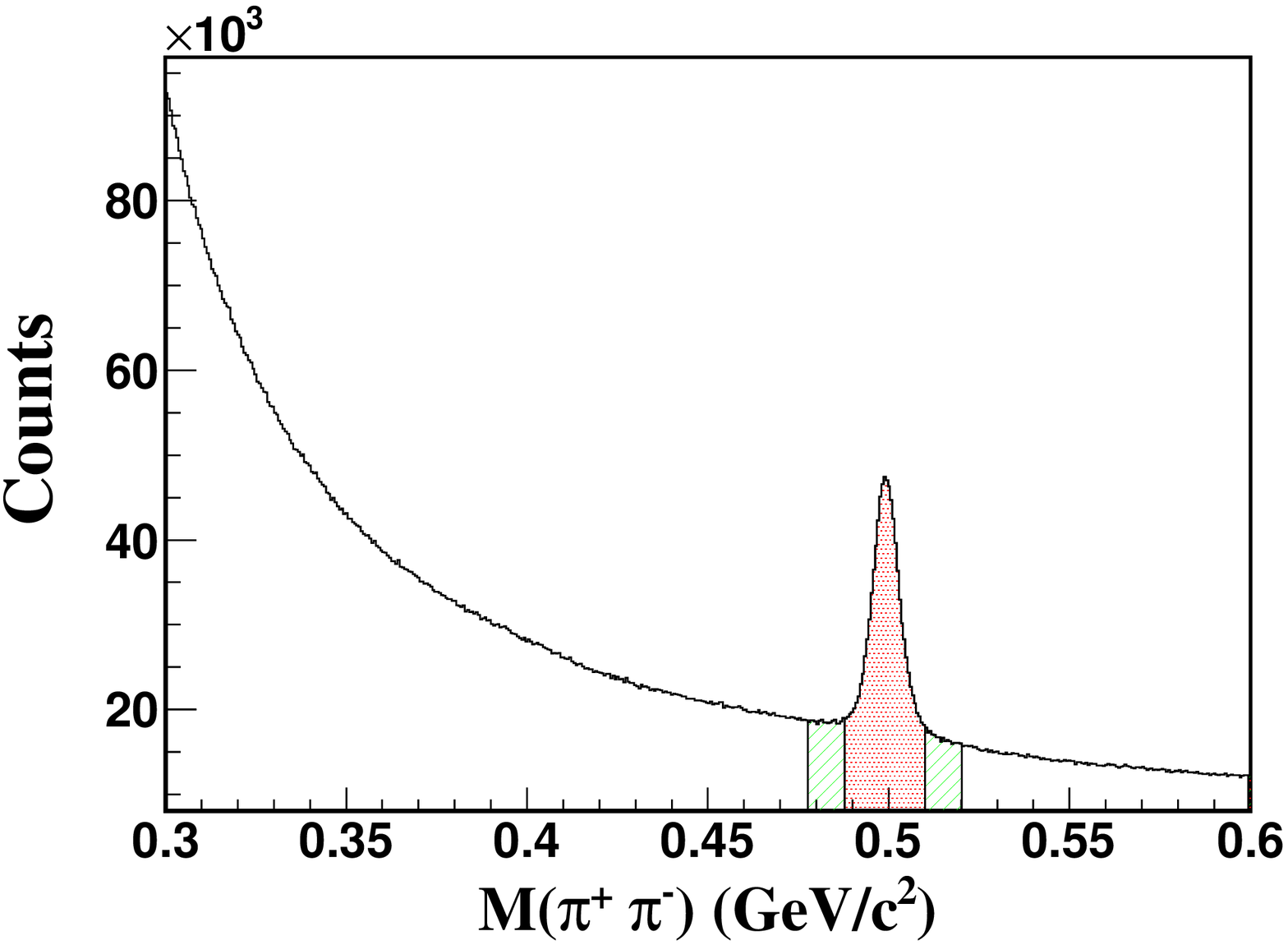}}
				\label{fig:m_PipPim}
			}
			\hspace{1.0cm}
			\subfigure[]
			{
				\resizebox{7.5cm}{!}{\includegraphics*{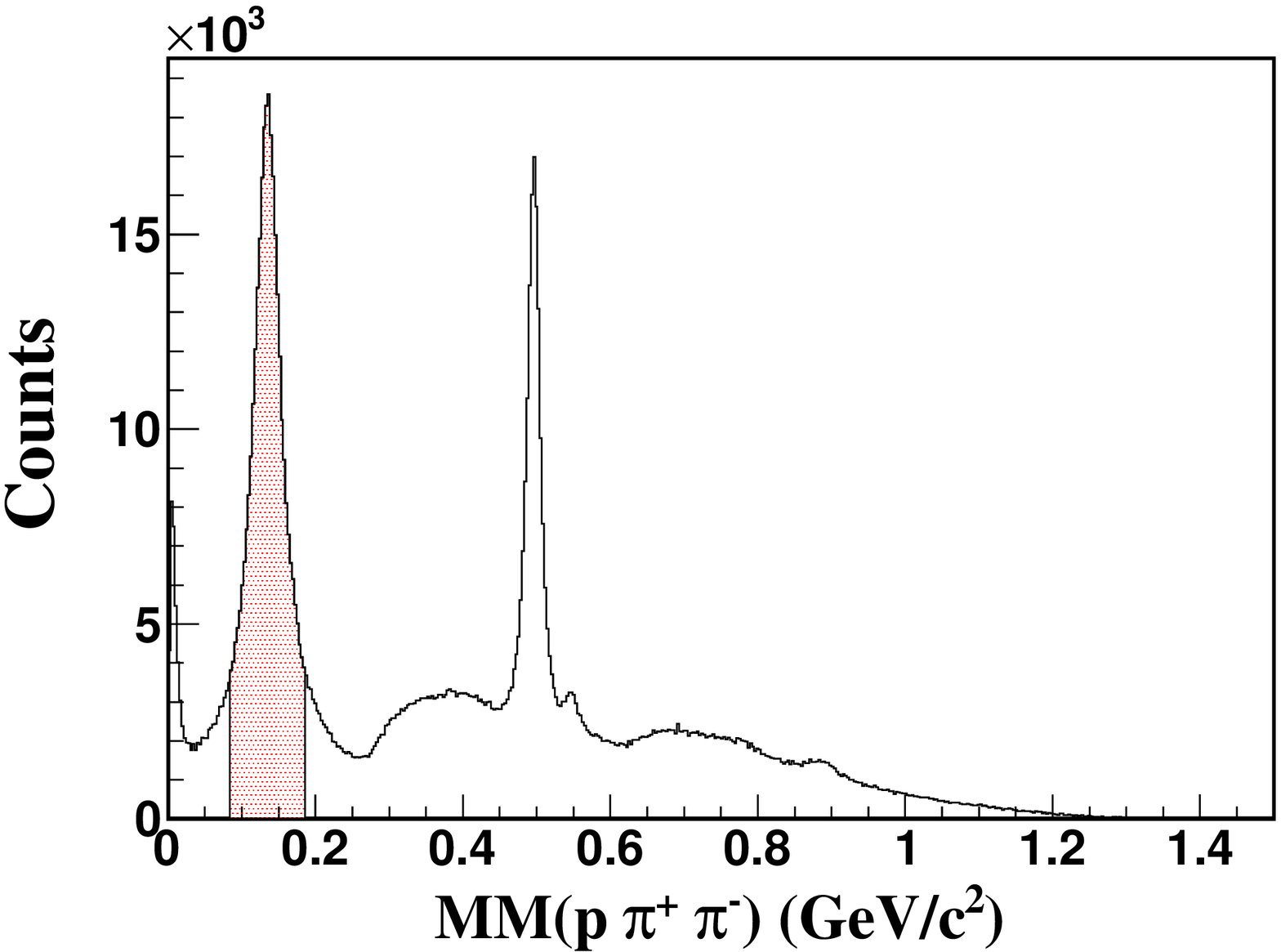}}
				\label{fig:mm_PPipPim}
			}
			\caption{(Color online) (a) Reconstructed $K^{0}_{S}$ signal in the invariant mass of the two charged pions with the collinearity angle cut. 
				The shaded region shows
				the $\pm3\sigma$ cut around the peak to select the $K^{0}_{S}$ and the striped region shows the $3\sigma$ sideband region of the $K^{0}_{S}$ taken
				to determine the background under the $\Sigma^{+}$ (see Sec.~\ref{background} for details),
				(b) Reconstructed $\pi^{0}$ signal in the missing mass of the detected proton and two oppositely charged pions with a $K^{0}_{S}$ 
				selected and the collinearity angle cut. The shaded region shows the $\pm3\sigma$ cut around the peak to select the $\pi^{0}$ events.}
		\end{center}
	\end{figure*}

\begin{figure}[hbt]
		\begin{center}
			\resizebox{7.5cm}{!}{\includegraphics*{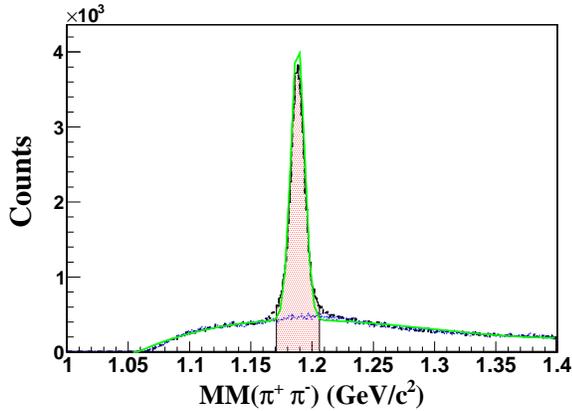}}
			\caption{(Color online) Reconstructed $\Sigma^{+}$ signal (dashed line) in the missing mass of the two charged pions, with the $\pi^{0}$ and 
			the $K^{0}_{S}$ selected, and with the collinearity angle cut as described in the text. 
			The dotted line is the $MM(\pi^{+}\pi^{-})$ distribution from the sidebands of the $K^{0}_{S}$ with a $\pi^{0}$ selected 
			and with the same collinearity angle cut as above. 
			No normalization or scaling is done. 
			The fit (solid line) includes a Gaussian for the peak and a second order polynomial for the background.
			The shaded region shows the $\pm3\sigma$ cut around the peak to select the $\Sigma^{+}$.} 
			\label{fig:mm_PipPim}
		\end{center}
	\end{figure}

	The reaction in Eq.~\ref{equ:channel} was reconstructed in the following way: the $\pi^{0}$ was reconstructed from the missing mass of the proton and two 
	oppositely charged pions, the $K^{0}_{S}$ was reconstructed from the invariant 
	mass of the two oppositely charged pions, and the $\Sigma^{+}$ was reconstructed from the missing mass of the $K^{0}_{S}$.
	The following cuts were applied to the data:
	\begin{itemize}
		\item{The momentum direction of the reconstructed $K^{0}_{S}$ should be along the line joining the center of the distance of closest approach 
			(DOCA) of the two
			charged pions and the center of the distance of closest approach (DOCA) of the proton and the photon. We applied a cut
			$\cos{\theta}_{\text{collinearity}} > 0.98$ on this mismatch angle, called here the collinearity cut, as shown in Fig.~\ref{fig:dist}.
			The cosine of the collinearity angle distribution after cuts to select the $\pi^{0}$ and the $K^{0}_{S}$ is shown in Fig.~\ref{fig:coll}.}
		\item{A cut on the invariant mass of the two charged pions to select the $K^{0}_{S}$, $|M(\pi^{+}\pi^{-})-M_{K_{S}}| < 3\sigma$, where
			$M_{K_{S}}$ and $\sigma$ are the fitted values of mass and width of the $K_{S}$, respectively, from the $M(\pi^{+}\pi^{-})$ distribution.
			See Table~\ref{table:table1} and Fig.~\ref{fig:m_PipPim}.}
		\item{A cut on the missing mass of the proton and two charged pions to select the $\pi^{0}$, $|MM(p\pi^{+}\pi^{-})-M_{\pi^{0}}| < 3\sigma$, 
			where $M_{\pi^{0}}$ and $\sigma$ are the fitted values of mass and width of the $\pi^{0}$, respectively, from the 
			$MM(p\pi^{+}\pi^{-})$ distribution. See Table~\ref{table:table1} and Fig.~\ref{fig:mm_PPipPim}.}
		\item{A cut on the missing mass of the two charged pions to select $\Sigma^{+}$, $|MM(\pi^{+}\pi^{-})-M_{\Sigma^{+}}| < 3\sigma$,
			where $M_{\Sigma^{+}}$ and $\sigma$ are the fitted values of mass and width of the $\Sigma^{+}$, respectively, from the $MM(\pi^{+}\pi^{-})$
			distribution. See Table~\ref{table:table1} and Fig.~\ref{fig:mm_PipPim}.}
	\end{itemize}
	Figs.~\ref{fig:m_PipPim}, ~\ref{fig:mm_PPipPim} and ~\ref{fig:mm_PipPim} show the reconstructed $K^{0}_{S}$, $\pi^{0}$, and $\Sigma^{+}$ respectively.
	The  fitted values of the mass and Gaussian width of the $K^{0}_{S}$, $\pi^{0}$ and $\Sigma^{+}$ are shown in Table~\ref{table:table1}. 
	The $\Sigma^{+}$ is selected, for final calculation, by taking a cut $|MM(\pi^{+}\pi^{-})-M_{\Sigma^{+}}| < 3\sigma$ in addition to     
	the above mentioned cuts. Here, $M_{\Sigma^{+}}$ and $\sigma$ are the fitted values of mass and width of the $\Sigma^{+}$. See Table~\ref{table:table1}.
	\begin{table}[hbt]
		\caption{Fitted values of the mass and Gaussian width of the different reconstructed particles. The fitting is done with a Gaussian (for the peak)
		+ $2^{nd}$ order polynomial  (for the background) function.}
	\begin{tabular}{|l|c|c|p{2.5cm}|}
		\hline
		particle        &       mass (GeV/c$^{2})$      & width, $\sigma$ (GeV/c$^{2})$   & cuts applied   \\
		\hline
		$K^{0}_{S}$	&       0.4990                  & 0.0036                & collinearity cut \\
		$\pi^{0}$       &       0.1351                  & 0.0169                & collinearity cut and $K^{0}_{S}$ selected \\
		$\Sigma^{+}$    &       1.1883                  & 0.0056                & collinearity cut, $K_{S}$ and $\pi^{0}$ selected \\
		\hline
	\end{tabular}
	\label{table:table1}
	\end{table}
\section{Background subtraction}
\label{background}
	One of the sources of physics background is from $\omega$ production. Because $\omega$ decays to $\pi^{+}\pi^{-}\pi^{0}$, 
	it is also present in the sidebands of the $K^{0}_{S}$. There is also a background due to direct production of the final state particles. 
	The dotted line in 
	Fig.~\ref{fig:mm_PipPim} shows the missing mass distribution of the two oppositely charged pions after selecting $3\sigma$ wide sidebands from  
	both sides of the $K^{0}_{S}$ in the $M(\pi^{+}\pi^{-})$ distribution. Here, we take $|M(\pi^{+}\pi^{-})-M_{K_{S}}+4.5\sigma|<1.5\sigma$
	for the left sideband and $|M(\pi^{+}\pi^{-})-M_{K_{S}}-4.5\sigma|<1.5\sigma$ for the right sideband, where $M_{K_{S}}$ and $\sigma$ are the fitted values of the mass and 
	Gaussian width of the $K^{0}_{S}$ (see Table~\ref{table:table1}). No normalization or scaling was applied. As we can see from Fig.~\ref{fig:mm_PipPim}, 
	the background is perfectly described by the sidebands of the $K^{0}_{S}$. We also checked the sideband distributions for the different kinematic bins 
	used in the final results, and we found that the sidebands perfectly describe the background in all kinematic bins. Therefore, we used the sidebands of the 
	$K^{0}_{S}$ under the $\Sigma^{+}$ peak for the background subtraction. We checked the background due to misidentification of kaons as protons and 
	it was negligible. 
\section{Detector acceptance correction}
\label{accp}
	The Monte Carlo (MC) events were generated uniformly in the ($K^{0}_{S} \Sigma^+$) phase space with a uniform angular distribution of the proton in 
	the $\Sigma^{+}$ rest frame, {\it i.e.} with zero polarization. The CLAS GEANT based simulation tool was used to simulate the passage of the generated events through CLAS. 
	Then, the accepted events were reconstructed by using the CLAS reconstruction software. 
	Distributions of different kinematic variables from the accepted MC events were compared with the experimental data and showed good 
	agreement. The up and down acceptance distributions with respect to the production plane were equal to within less than 1\%. See Sec.~\ref{results} for 
	the definition of the up and down distributions. The polarization calculated from the accepted events was less than $2\%$ in the entire kinematic range of 
	our measurement. Therefore, the effects due to detector acceptance and false asymmetry are negligible. These MC events were used for the acceptance correction.

	To check the quality of the acceptance correction, we also generated MC events with 100\% polarization by using the same MC generator
	and reconstructed by the CLAS reconstruction software. We applied the acceptance correction to the accepted events by using the acceptance 
	function obtained from unpolarized MC events as explained above. Then, we calculated the polarization of the acceptance corrected  
	events and found that it is close to 100\% within $\pm2\%$. See Section~\ref{results} for a discussion of the polarization calculation method. From these studies 
	we concluded that our acceptance correction method works well, and the overall systematic uncertainty on the observed polarization due to the detector 
	acceptance and bias is $\approx$ 2\%. 
\section{Analysis Method and Results}
\label{results}
\begin{figure}[hbt]
		\begin{center}
			\resizebox{7cm}{!}{\includegraphics*{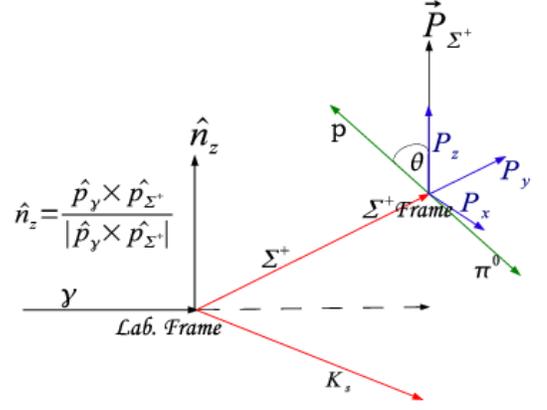}}
			\caption{(Color online) Coordinate system.}
			\label{fig:axis}
		\end{center}
	\end{figure}

	The $\Sigma^{+}$ is produced via the electromagnetic interaction, which conserves parity. However, it decays to a proton and $\pi^{0}$ via
	the parity violating weak interaction. Therefore, the polarization of the $\Sigma^{+}$ can be measured from the angular distribution of
	one of its decay products in the $\Sigma^{+}$ rest frame. Below we take the direction normal to the production plane as the $z$-axis (transversity 
	frame~\cite{Beaupre:1973re}), the direction along the $\Sigma^{+}$ momentum vector as the $y$-axis, and the $x$-axis is chosen in order to make  
	a right-handed coordinate system, as shown in Fig.~\ref{fig:axis}. Corresponding unit-vectors are given by
	\begin{align}
		\hat{n}_{z} &~=~ \frac{\hat{p}_{\gamma} \times \hat{p}_{\Sigma^+}}{|\hat{p}_{\gamma} \times \hat{p}_{\Sigma^+}|}, \\
		\hat{n}_{y} &~=~ \hat{p}_{\Sigma^+}, \\
		\hat{n}_{x} &~=~ \hat{n}_{y} \times \hat{n}_{z}.
	\end{align}

	The angular distribution of the proton in the $\Sigma^{+}$ rest frame is given by~\cite{Cronin:1963zb}
	\begin{eqnarray}
		\frac{dN}{d\cos\theta} ~=~ \frac{N_{0}}{4\pi} ~[1 ~+~ \alpha_{\Sigma^{+}} ~P_{\Sigma^{+}} \cos\theta],
		\label{eq:dist}
	\end{eqnarray}
	where $\theta$ is the angle between the proton momentum vector and the quantization axis of the $\Sigma^{+}$ along $\hat{n}_{z}$, $P_{\Sigma^{+}}$
	is the transverse component of the polarization of the $\Sigma^{+}$, $\alpha_{\Sigma^{+}}$ is a measure of the degree of parity mixing~\cite{Lee:1957qs} and
	its value for the above decay channel is $-0.980^{+0.017}_{-0.015}$~\cite{pdg}. $N_{0}$ is the total number of events.
	The longitudinal component of the polarization vanishes. Eq.~\eqref{eq:dist}
	can be split into up ($N^{U}$) and down ($N^{D}$) distributions with respect to the production plane:
	\begin{align}
		\label{equ:one}
		N^{U}(\cos\theta) &= \frac{dN^{U}}{d\cos\theta} \nonumber \\ 
				  &= \frac{N_{0}}{4\pi} [1 + \alpha_{\Sigma^{+}} P_{\Sigma^{+}} \cos\theta] ~\text{for}~ 0 \leq \cos\theta \leq 1,  \\
		N^{D}(\cos\theta) &= \frac{dN^{D}}{d\cos\theta} \nonumber \\
				  &= \frac{N_{0}}{4\pi} [1 - \alpha_{\Sigma^{+}} P_{\Sigma^{+}} \cos\theta] ~\text{for}~ -1 \leq \cos\theta \leq 0.
		\label{equ:two}
	\end{align}
	Using these two equations, one can write
	\begin{eqnarray}
		\alpha_{\Sigma^{+}} P_{\Sigma^{+}}\cos\theta = \Big[\frac{N^{U}(\cos\theta) ~-~ N^{D}(\cos\theta)}{N^{U}(\cos\theta) ~+~ N^{D}(\cos\theta)}\Big].
		\label{equ:final1}
	\end{eqnarray}
	Here, $\cos\theta$ varies from 0 to 1 only. The benefit of using the ratio of the up and down distributions is essentially to cancel the effect of
	the acceptance correction, assuming that the acceptance corrections for the up and down distributions are the same. However, we didn't rely on such an
	assumption and applied acceptance corrections. Eq.~\eqref{equ:final1} can be integrated over $\cos\theta$ to obtain
	\begin{eqnarray}
		P_{\Sigma^{+}} = \frac{2}{\alpha_{\Sigma^{+}}} \Big[\frac{N^{U} ~-~ N^{D}}{N^{U} ~+~ N^{D}}\Big].
		\label{equ:final2}
	\end{eqnarray}
	In Eq.~\eqref{equ:final2}, the distributions are corrected bin-by-bin in $\cos(\theta_{\Sigma^{+}})_{CM}$ and the photon energy 
	for the CLAS acceptance.

	\begin{figure*}[hbt]
		\begin{center}
			\resizebox{18cm}{!}{\includegraphics*{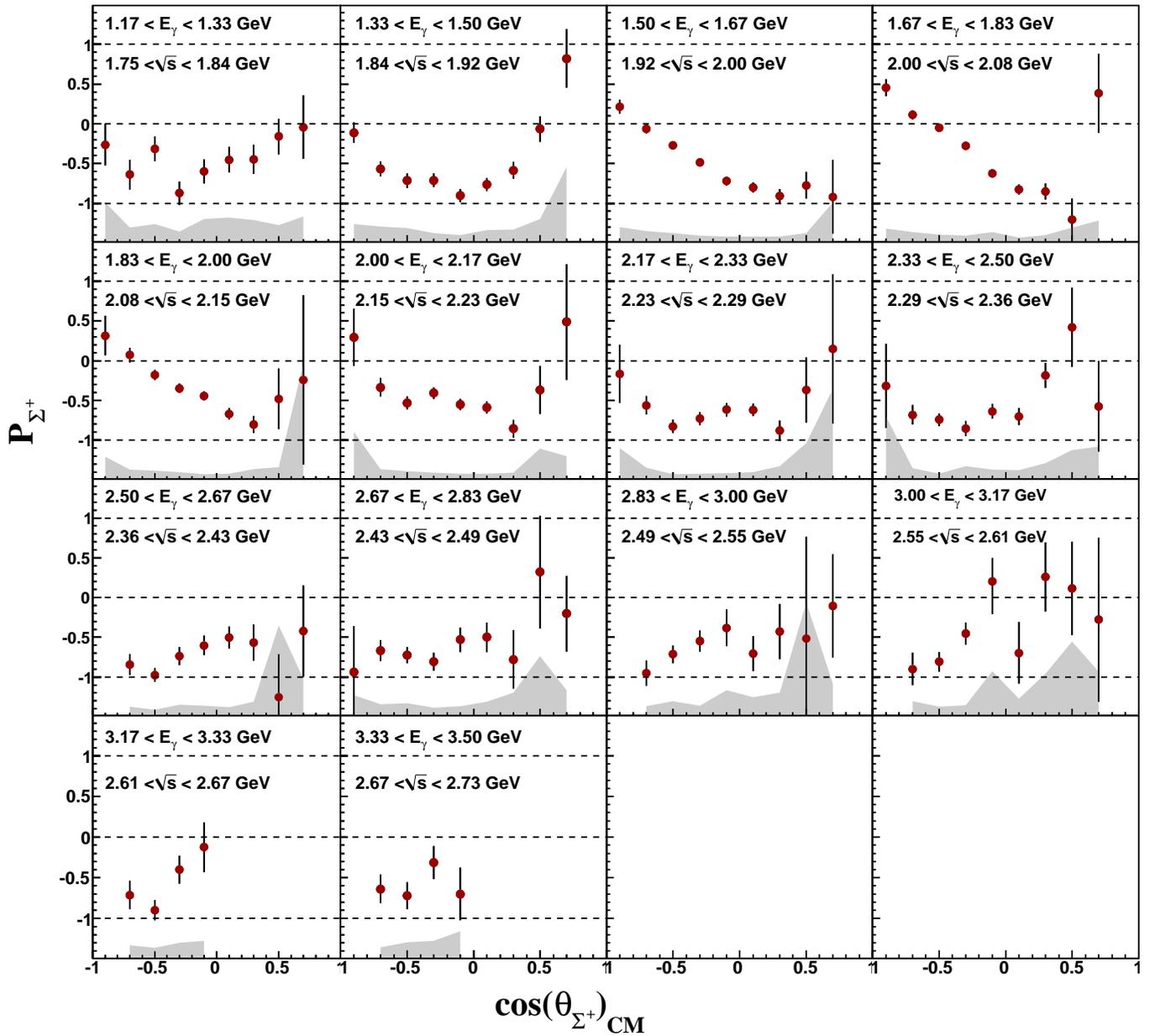}}
			\caption{(Color online) Transverse polarization versus 
				$\cos(\theta_{\Sigma^{+}})_{\text{CM}}$ 
				at different photon beam energies. The bands on the horizontal axis are the systematic uncertainties.}
			\label{fig:polCosDivideAccpCorrBkSubt}
		\end{center}
	\end{figure*}
	\begin{figure*}[hbt]
		\begin{center}
			\resizebox{16cm}{!}{\includegraphics*{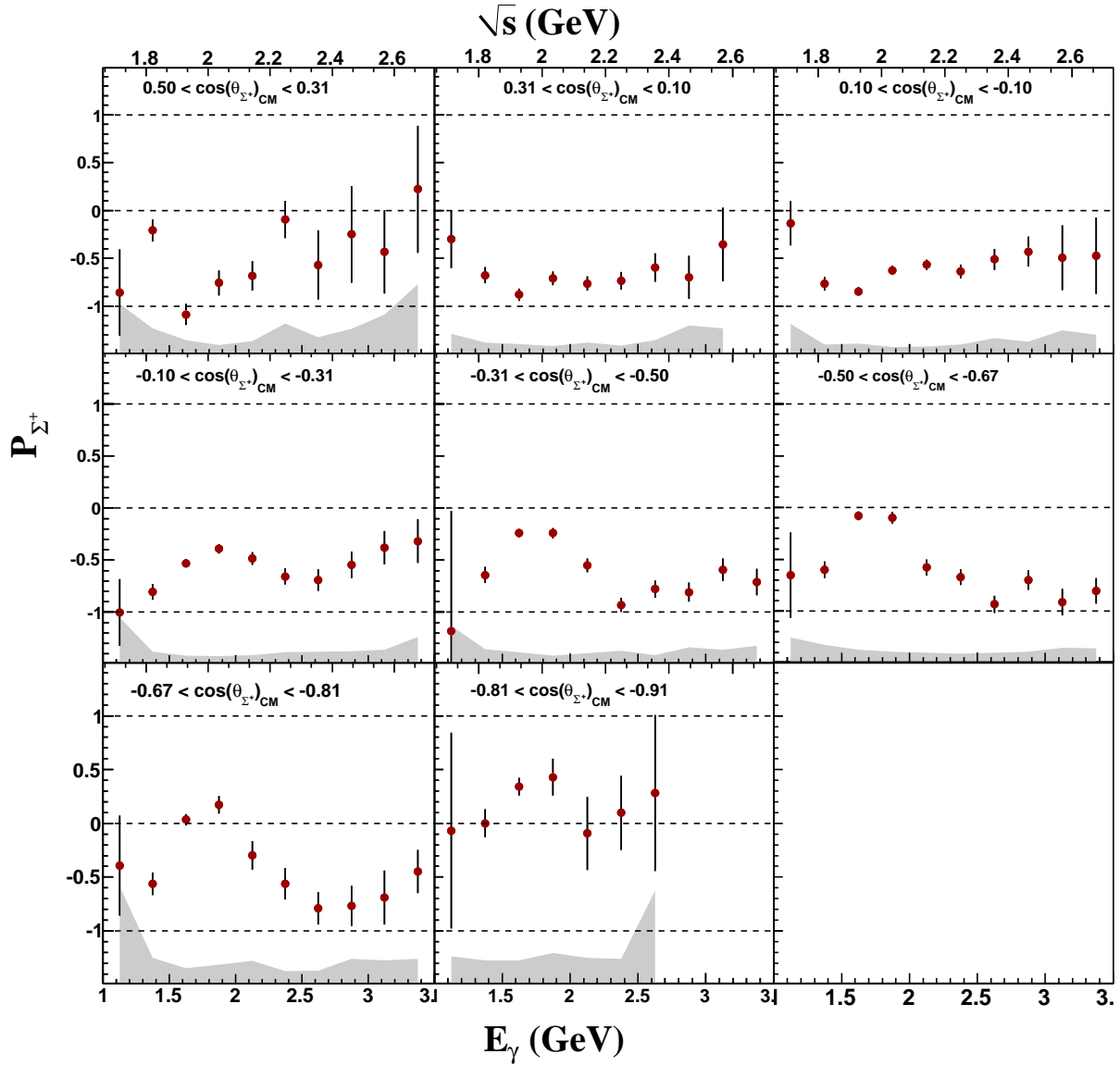}}
			\caption{(Color online) Transverse polarization versus photon beam energy at different $(\theta_{\Sigma^{+}})_{\text{CM}}$.
			The bands on the horizontal axis are the systematic uncertainties.}
			\label{fig:polEgDivideAccpCorrBkSubt}
		\end{center}
	\end{figure*}
	Fig.~\ref{fig:polCosDivideAccpCorrBkSubt} shows the polarization with respect to the production angle of $\Sigma^{+}$ in the $\gamma p$ CM frame,
	$(\theta_{\Sigma^{+}})_{CM}$,
	for different bins of photon energy from 1.0 GeV to 3.5 GeV.
	Fig.~\ref{fig:polEgDivideAccpCorrBkSubt} shows the polarization with respect to the photon energy for different $(\theta_{\Sigma^{+}})_{CM}$ bins.
	The error bars on the points are statistical uncertainties, the bands on the horizontal axis are the systematic uncertainties. The data points corresponding to
	Figs.~\ref{fig:polCosDivideAccpCorrBkSubt} and ~\ref{fig:polEgDivideAccpCorrBkSubt} are shown in Tables~\ref{table:table2}, ~\ref{table:table3} and ~\ref{table:table4}.
\section{Systematic Uncertainties}
\label{uncertainty}
	Systematic uncertainties are estimated from four different sources: 
	\begin{itemize}
		\item{mass cut: we changed the width for the $\Sigma^{+}$ selection from $\pm3\sigma$ to $\pm4\sigma$ and the difference in 
			polarizations obtained from these two selections is taken as a systematic uncertainty. The systematic uncertainty due to the mass cut varies up-to $\pm0.10$ 
			in most of the kinematic region.}
		\item{collinearity cut: we changed the collinearity cut from $\cos{\theta}_{\text{collinearity}} > 0.98$ to 
			$\cos{\theta}_{\text{collinearity}} > 0.90$ and the difference in polarizations obtained from these two cuts is taken as a systematic uncertainty. 
			The systematic uncertainty due to the collinearity cut varies up-to $\pm0.15$ in most of the kinematic region.}
		\item{background subtraction: polarization calculated from the background events is taken as a systematic uncertainty. An explanation of the background 
			events is given in Sec.~\ref{background}. The systematic uncertainty due to the background subtraction varies up-to $\pm0.05$ in most of 
			the kinematic region.}
		\item{acceptance correction: we did acceptance corrections to the data by using unpolarized MC events and 100\% polarized MC events separately, 
			and the difference in polarizations obtained from these two acceptance correction methods is taken as a systematic uncertainty. For the final 
			polarization results, the unpolarized MC events were used for the acceptance corrections to the data, see Sec.~\ref{accp}. The systematic uncertainty due to 
			the acceptance correction varies up-to $\pm0.05$ in most of the kinematic region.}
	
	\end{itemize}
	The most significant contribution to the systematic uncertainty comes from the collinearity cut. 
	The total systematic uncertainty for each bin is obtained by adding these four systematic uncertainties in that bin in quadrature and are shown by the grey 
	bands on Figs.~\ref{fig:polCosDivideAccpCorrBkSubt} and 
	~\ref{fig:polEgDivideAccpCorrBkSubt}. Errors are also shown in the Tables~\ref{table:table2},~\ref{table:table3}, and ~\ref{table:table4},
	along with the polarization values, where superscripts are statistical uncertainties and subscripts are systematic uncertainties.
	\begin{center}
	\begin{table*}[ht!]
	\caption{Bin averaged polarization vs $\cos(\theta_{\Sigma^{+}})_{\text{CM}}$ for different $E_{\gamma}$ and corresponding $\sqrt{s}$ bins. The first line and the second line
	in the second row in the right columns are the $E_{\gamma}$ ranges and the corresponding $\sqrt{s}$ ranges respectively. Superscripts are statistical errors and subscripts
	are systematic errors. The data points are taken from Fig.~\ref{fig:polCosDivideAccpCorrBkSubt}.}
	\begin{tabular}{|c|c|c|c|c|c|c|c|}
	\hline
	$\cos(\theta_{\Sigma^{+}})_{\text{CM}}$ & \multicolumn{7}{|c|}{$E_{\gamma}$ (GeV) / $\sqrt{s}$ (GeV)} \\
	\cline{2-8}
	&  \begin{tabular}[c]{@{}c@{}}$1.17$ - $1.33$ \\$1.75$ - $1.84$ \end{tabular} &  \begin{tabular}[c]{@{}c@{}}$1.33$ - $1.50$ \\$1.84$ - $1.92$ \end{tabular} &  \begin{tabular}[c]{@{}c@{}}$1.50$ - $1.67$ \\$1.92$ - $2.00$ \end{tabular} &  \begin{tabular}[c]{@{}c@{}}$1.67$ - $1.83$ \\$2.00$ - $2.08$ \end{tabular} &  \begin{tabular}[c]{@{}c@{}}$1.83$ - $2.00$ \\$2.08$ - $2.15$ \end{tabular} &  \begin{tabular}[c]{@{}c@{}}$2.00$ - $2.17$ \\$2.15$ - $2.23$ \end{tabular} &  \begin{tabular}[c]{@{}c@{}}$2.17$ - $2.33$\\$2.23$ - $2.29$ \end{tabular} \\
        \hline
        $-1.00$ - $-0.80$  &  $-0.26^{\pm0.26}_{\pm0.51}$  &  $-0.11^{\pm0.13}_{\pm0.24}$  &  $0.21^{\pm0.09}_{\pm0.20}$  &  $0.45^{\pm0.11}_{\pm0.18}$  &  $0.31^{\pm0.25}_{\pm0.29}$  &  $0.29^{\pm0.36}_{\pm0.60}$  &  $-0.17^{\pm0.37}_{\pm0.40}$  \\
        $-0.80$ - $-0.60$  &  $-0.64^{\pm0.19}_{\pm0.19}$  &  $-0.57^{\pm0.09}_{\pm0.21}$  &  $-0.06^{\pm0.05}_{\pm0.15}$  &  $0.11^{\pm0.06}_{\pm0.14}$  &  $0.07^{\pm0.09}_{\pm0.13}$  &  $-0.33^{\pm0.11}_{\pm0.13}$  &  $-0.56^{\pm0.12}_{\pm0.16}$  \\
        $-0.60$ - $-0.40$  &  $-0.31^{\pm0.16}_{\pm0.24}$  &  $-0.71^{\pm0.09}_{\pm0.19}$  &  $-0.27^{\pm0.05}_{\pm0.13}$  &  $-0.05^{\pm0.05}_{\pm0.11}$  &  $-0.18^{\pm0.06}_{\pm0.12}$  &  $-0.53^{\pm0.08}_{\pm0.11}$  &  $-0.83^{\pm0.08}_{\pm0.07}$  \\
        $-0.40$ - $-0.20$  &  $-0.87^{\pm0.15}_{\pm0.14}$  &  $-0.71^{\pm0.08}_{\pm0.13}$  &  $-0.48^{\pm0.05}_{\pm0.09}$  &  $-0.28^{\pm0.05}_{\pm0.09}$  &  $-0.35^{\pm0.06}_{\pm0.10}$  &  $-0.41^{\pm0.07}_{\pm0.09}$  &  $-0.73^{\pm0.08}_{\pm0.08}$  \\
        $-0.20$ - $0.00$  &  $-0.60^{\pm0.15}_{\pm0.30}$  &  $-0.90^{\pm0.08}_{\pm0.10}$  &  $-0.72^{\pm0.05}_{\pm0.08}$  &  $-0.62^{\pm0.05}_{\pm0.14}$  &  $-0.44^{\pm0.06}_{\pm0.07}$  &  $-0.55^{\pm0.07}_{\pm0.08}$  &  $-0.62^{\pm0.08}_{\pm0.08}$  \\
        $0.00$ - $0.20$  &  $-0.45^{\pm0.16}_{\pm0.32}$  &  $-0.76^{\pm0.08}_{\pm0.16}$  &  $-0.80^{\pm0.07}_{\pm0.08}$  &  $-0.83^{\pm0.06}_{\pm0.07}$  &  $-0.67^{\pm0.07}_{\pm0.08}$  &  $-0.59^{\pm0.07}_{\pm0.08}$  &  $-0.62^{\pm0.08}_{\pm0.10}$  \\
        $0.20$ - $0.40$  &  $-0.45^{\pm0.18}_{\pm0.29}$  &  $-0.59^{\pm0.11}_{\pm0.17}$  &  $-0.91^{\pm0.09}_{\pm0.08}$  &  $-0.85^{\pm0.10}_{\pm0.10}$  &  $-0.80^{\pm0.11}_{\pm0.13}$  &  $-0.86^{\pm0.11}_{\pm0.09}$  &  $-0.88^{\pm0.13}_{\pm0.18}$  \\
        $0.40$ - $0.60$  &  $-0.16^{\pm0.23}_{\pm0.23}$  &  $-0.07^{\pm0.16}_{\pm0.30}$  &  $-0.77^{\pm0.17}_{\pm0.12}$  &  $-1.21^{\pm0.26}_{\pm0.19}$  &  $-0.48^{\pm0.38}_{\pm0.16}$  &  $-0.37^{\pm0.30}_{\pm0.40}$  &  $-0.37^{\pm0.41}_{\pm0.46}$  \\
        $0.60$ - $0.80$  &  $-0.04^{\pm0.40}_{\pm0.33}$  &  $0.82^{\pm0.37}_{\pm0.96}$  &  $-0.92^{\pm0.46}_{\pm0.51}$  &  $0.38^{\pm0.50}_{\pm0.28}$  &  $-0.24^{\pm1.07}_{\pm1.47}$  &  $0.49^{\pm0.73}_{\pm0.30}$  &  $0.15^{\pm0.94}_{\pm1.14}$  \\
        \hline
	\end{tabular}
	\label{table:table2}
	\end{table*}
	\end{center}
	\begin{center}
	\begin{table*}[ht!]
	\caption{Bin averaged polarization vs $\cos(\theta_{\Sigma^{+}})_{\text{CM}}$ for different $E_{\gamma}$ and corresponding $\sqrt{s}$ bins. Thef first line and the second line
	in the second row in the right columns are the $E_{\gamma}$ ranges and the corresponding $\sqrt{s}$ ranges respectively. Superscripts are statistical errors and subscripts
	are systematic errors. The data points are taken from Fig.~\ref{fig:polCosDivideAccpCorrBkSubt}.}
	\begin{tabular}{|c|c|c|c|c|c|c|c|}
	\hline
	$\cos(\theta_{\Sigma^{+}})_{\text{CM}}$ & \multicolumn{7}{|c|}{$E_{\gamma}$ (GeV) / $\sqrt{s}$ (GeV)} \\
	\cline{2-8}
	&  \begin{tabular}[c]{@{}c@{}}$2.33$ - $2.50$ \\$2.29$ - $2.36$ \end{tabular} &  \begin{tabular}[c]{@{}c@{}}$2.50$ - $2.67$ \\$2.36$ - $2.43$ \end{tabular} &  \begin{tabular}[c]{@{}c@{}}$2.67$ - $2.83$ \\$2.43$ - $2.49$ \end{tabular} &  \begin{tabular}[c]{@{}c@{}}$2.83$ - $3.00$ \\$2.49$ - $2.55$ \end{tabular} &  \begin{tabular}[c]{@{}c@{}}$3.00$ - $3.17$ \\$2.55$ - $2.61$ \end{tabular} &  \begin{tabular}[c]{@{}c@{}}$3.17$ - $3.33$ \\$2.61$ - $2.67$ \end{tabular} &  \begin{tabular}[c]{@{}c@{}}$3.33$ - $3.50$\\$2.67$ - $2.73$ \end{tabular} \\
        \hline
        $-1.00$ - $-0.80$  &  $-0.32^{\pm0.53}_{\pm0.80}$  &  $-0.84^{\pm0.13}_{\pm0.13}$  &  $-0.94^{\pm0.58}_{\pm0.27}$  &  $-0.95^{\pm0.16}_{\pm0.13}$  &  $-0.90^{\pm0.21}_{\pm0.19}$  &  $-0.71^{\pm0.18}_{\pm0.17}$  &  $-0.64^{\pm0.17}_{\pm0.14}$  \\
        $-0.80$ - $-0.60$  &  $-0.68^{\pm0.12}_{\pm0.15}$  &  $-0.97^{\pm0.08}_{\pm0.09}$  &  $-0.67^{\pm0.13}_{\pm0.16}$  &  $-0.71^{\pm0.11}_{\pm0.19}$  &  $-0.81^{\pm0.12}_{\pm0.13}$  &  $-0.90^{\pm0.12}_{\pm0.14}$  &  $-0.72^{\pm0.17}_{\pm0.20}$  \\
        $-0.60$ - $-0.40$  &  $-0.74^{\pm0.08}_{\pm0.08}$  &  $-0.74^{\pm0.11}_{\pm0.15}$  &  $-0.72^{\pm0.10}_{\pm0.17}$  &  $-0.55^{\pm0.13}_{\pm0.14}$  &  $-0.45^{\pm0.14}_{\pm0.15}$  &  $-0.40^{\pm0.17}_{\pm0.20}$  &  $-0.31^{\pm0.20}_{\pm0.22}$  \\
        $-0.40$ - $-0.20$  &  $-0.85^{\pm0.09}_{\pm0.18}$  &  $-0.60^{\pm0.13}_{\pm0.14}$  &  $-0.81^{\pm0.11}_{\pm0.11}$  &  $-0.38^{\pm0.23}_{\pm0.34}$  &  $0.20^{\pm0.41}_{\pm0.58}$  &  $-0.12^{\pm0.31}_{\pm0.23}$  &  $-0.70^{\pm0.32}_{\pm0.34}$  \\
        $-0.20$ - $0.00$  &  $-0.64^{\pm0.09}_{\pm0.13}$  &  $-0.50^{\pm0.14}_{\pm0.12}$  &  $-0.53^{\pm0.15}_{\pm0.13}$  &  $-0.71^{\pm0.22}_{\pm0.25}$  &  $-0.70^{\pm0.39}_{\pm0.22}$  &  -  &  -  \\
        $0.00$ - $0.20$  &  $-0.70^{\pm0.11}_{\pm0.12}$  &  $-0.57^{\pm0.23}_{\pm0.19}$  &  $-0.50^{\pm0.19}_{\pm0.19}$  &  $-0.43^{\pm0.35}_{\pm0.30}$  &  $0.26^{\pm0.43}_{\pm0.54}$  &  -  &  -  \\
        $0.20$ - $0.40$  &  $-0.19^{\pm0.16}_{\pm0.21}$  &  $-1.25^{\pm0.54}_{\pm1.15}$  &  $-0.78^{\pm0.37}_{\pm0.30}$  &  $-0.51^{\pm1.28}_{\pm1.44}$  &  $0.11^{\pm0.59}_{\pm0.94}$  &  -  &  -  \\
        $0.40$ - $0.60$  &  $0.42^{\pm0.50}_{\pm0.37}$  &  $-0.42^{\pm0.57}_{\pm0.48}$  &  $0.32^{\pm0.71}_{\pm0.76}$  &  $-0.11^{\pm0.65}_{\pm0.42}$  &  $-0.28^{\pm1.03}_{\pm0.57}$  &  -  &  -  \\
        $0.60$ - $0.80$  &  $-0.57^{\pm0.57}_{\pm0.42}$  &  -  &  $-0.20^{\pm0.48}_{\pm0.33}$  &  -  &  -  &  -  &  -  \\
        \hline
	\end{tabular}
	\label{table:table3}
	\end{table*}
	\end{center}
	\begin{center}
	\begin{table*}[ht!]
	\caption{Bin averaged polarization vs $E_{\gamma}$ for different $\cos(\theta_{\Sigma^{+}})_{\text{CM}}$ bins. Superscripts are statistical errors and subscripts are systematic errors. The data points are taken from Fig.~\ref{fig:polEgDivideAccpCorrBkSubt}.}
	\begin{tabular}{|c|c|c|c|c|c|c|c|c|c|}
	\hline
	$E_{\gamma}$ (GeV) & $\sqrt{s}$ (GeV) & \multicolumn{8}{|c|}{$\cos(\theta_{\Sigma^{+}})_{\text{CM}}$} \\
	\cline{3-10}
         &  & $0.50$ - $0.31$  &  $0.31$ - $0.10$  &  $0.10$ - $-0.10$  &  $-0.10$ - $-0.31$  &  $-0.31$ - $-0.50$  &  $-0.50$ - $-0.67$  &  $-0.67$ - $-0.81$  &  $-0.81$ - $-0.91$\\  \hline
        $1.00$ - $1.25$  &      $1.66$ - $1.80$  &  $-0.85^{\pm0.45}_{\pm0.53}$  &  $-0.30^{\pm0.30}_{\pm0.22}$  &  $-0.13^{\pm0.23}_{\pm0.31}$  &  $-1.00^{\pm0.32}_{\pm0.45}$  &  $-1.18^{\pm1.15}_{\pm0.37}$  &  $-0.65^{\pm0.41}_{\pm0.24}$  &  $-0.39^{\pm0.47}_{\pm0.91}$  &  $-0.07^{\pm0.91}_{\pm0.26}$  \\
        $1.25$ - $1.50$  &      $1.80$ - $1.92$  &  $-0.21^{\pm0.12}_{\pm0.27}$  &  $-0.68^{\pm0.08}_{\pm0.12}$  &  $-0.76^{\pm0.07}_{\pm0.10}$  &  $-0.81^{\pm0.08}_{\pm0.12}$  &  $-0.64^{\pm0.08}_{\pm0.15}$  &  $-0.60^{\pm0.08}_{\pm0.17}$  &  $-0.56^{\pm0.11}_{\pm0.25}$  &  $0.00^{\pm0.13}_{\pm0.23}$  \\
        $1.50$ - $1.75$  &      $1.92$ - $2.04$  &  $-1.08^{\pm0.11}_{\pm0.15}$  &  $-0.88^{\pm0.06}_{\pm0.11}$  &  $-0.84^{\pm0.04}_{\pm0.11}$  &  $-0.53^{\pm0.04}_{\pm0.08}$  &  $-0.24^{\pm0.04}_{\pm0.12}$  &  $-0.08^{\pm0.04}_{\pm0.13}$  &  $0.03^{\pm0.05}_{\pm0.15}$  &  $0.34^{\pm0.08}_{\pm0.23}$  \\
        $1.75$ - $2.00$  &      $2.04$ - $2.15$  &  $-0.76^{\pm0.13}_{\pm0.10}$  &  $-0.71^{\pm0.07}_{\pm0.08}$  &  $-0.62^{\pm0.05}_{\pm0.08}$  &  $-0.39^{\pm0.04}_{\pm0.08}$  &  $-0.24^{\pm0.05}_{\pm0.08}$  &  $-0.10^{\pm0.06}_{\pm0.11}$  &  $0.17^{\pm0.08}_{\pm0.18}$  &  $0.43^{\pm0.17}_{\pm0.29}$  \\
        $2.00$ - $2.25$  &      $2.15$ - $2.26$  &  $-0.68^{\pm0.15}_{\pm0.14}$  &  $-0.76^{\pm0.07}_{\pm0.12}$  &  $-0.57^{\pm0.06}_{\pm0.08}$  &  $-0.49^{\pm0.06}_{\pm0.08}$  &  $-0.55^{\pm0.07}_{\pm0.10}$  &  $-0.58^{\pm0.08}_{\pm0.10}$  &  $-0.30^{\pm0.13}_{\pm0.22}$  &  $-0.09^{\pm0.34}_{\pm0.25}$  \\
        $2.25$ - $2.50$  &      $2.26$ - $2.36$  &  $-0.09^{\pm0.20}_{\pm0.32}$  &  $-0.74^{\pm0.09}_{\pm0.09}$  &  $-0.64^{\pm0.07}_{\pm0.10}$  &  $-0.66^{\pm0.08}_{\pm0.12}$  &  $-0.93^{\pm0.07}_{\pm0.13}$  &  $-0.67^{\pm0.08}_{\pm0.09}$  &  $-0.56^{\pm0.15}_{\pm0.12}$  &  $0.10^{\pm0.34}_{\pm0.24}$  \\
        $2.50$ - $2.75$  &      $2.36$ - $2.46$  &  $-0.57^{\pm0.36}_{\pm0.17}$  &  $-0.59^{\pm0.15}_{\pm0.14}$  &  $-0.51^{\pm0.11}_{\pm0.17}$  &  $-0.69^{\pm0.10}_{\pm0.12}$  &  $-0.78^{\pm0.08}_{\pm0.08}$  &  $-0.93^{\pm0.08}_{\pm0.10}$  &  $-0.79^{\pm0.15}_{\pm0.13}$  &  $0.28^{\pm0.73}_{\pm0.87}$  \\
        $2.75$ - $3.00$  &      $2.46$ - $2.55$  &  $-0.25^{\pm0.50}_{\pm0.26}$  &  $-0.70^{\pm0.23}_{\pm0.30}$  &  $-0.43^{\pm0.16}_{\pm0.13}$  &  $-0.55^{\pm0.13}_{\pm0.13}$  &  $-0.81^{\pm0.09}_{\pm0.16}$  &  $-0.70^{\pm0.09}_{\pm0.11}$  &  $-0.77^{\pm0.19}_{\pm0.24}$  &   -   \\
        $3.00$ - $3.25$  &      $2.55$ - $2.64$  &  $-0.43^{\pm0.44}_{\pm0.42}$  &  $-0.35^{\pm0.38}_{\pm0.27}$  &  $-0.49^{\pm0.34}_{\pm0.25}$  &  $-0.38^{\pm0.16}_{\pm0.14}$  &  $-0.59^{\pm0.11}_{\pm0.14}$  &  $-0.91^{\pm0.13}_{\pm0.15}$  &  $-0.69^{\pm0.25}_{\pm0.23}$  &   -   \\
        $3.25$ - $3.50$  &      $2.64$ - $2.73$  &  $0.22^{\pm0.66}_{\pm0.73}$  &  - & $-0.47^{\pm0.40}_{\pm0.20}$  &  $-0.32^{\pm0.21}_{\pm0.26}$  &  $-0.71^{\pm0.13}_{\pm0.17}$  &  $-0.80^{\pm0.12}_{\pm0.14}$  &  $-0.45^{\pm0.20}_{\pm0.24}$  &   -   \\
        \hline
	\end{tabular}
	\label{table:table4}
	\end{table*}
	\end{center}
\section{Discussion and Summary}
\begin{figure*}[hbt]
		\begin{center}
			\resizebox{18cm}{!}{\includegraphics*{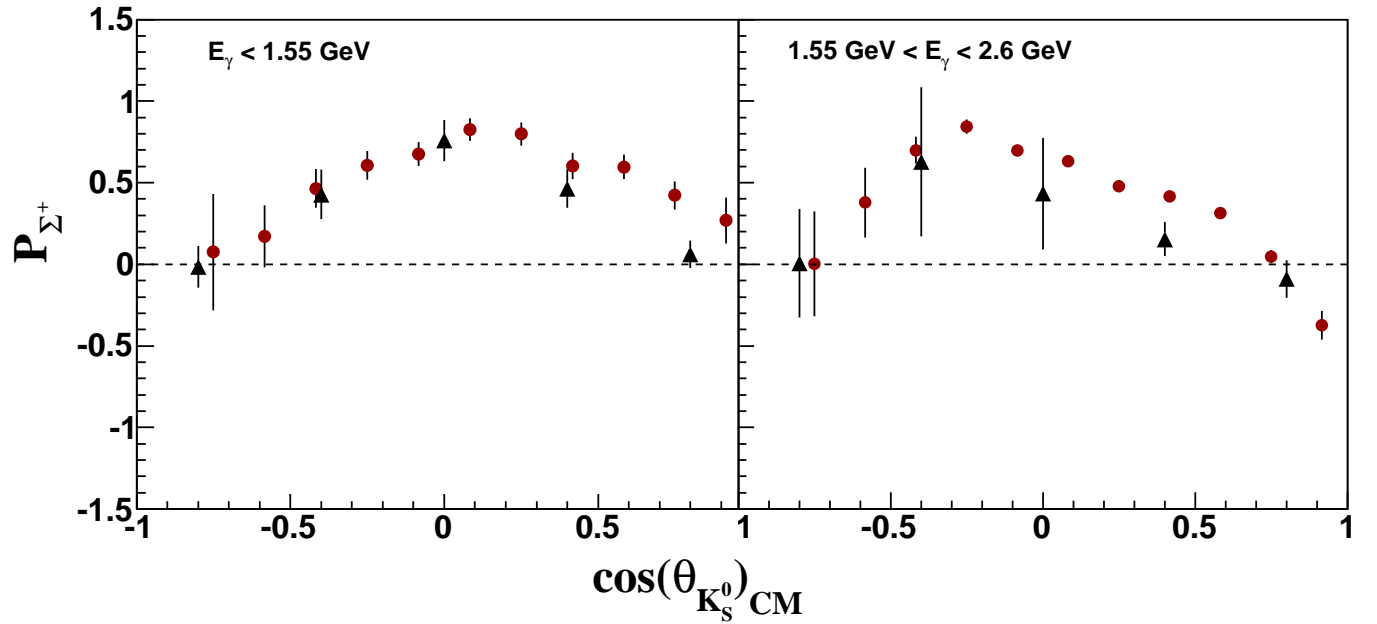}}
			\caption[Comparison of polarization of $\Sigma^{+}$ between (this result) and SAPHIR]{(Color online) Comparison of polarization of 
				$\Sigma^{+}$ between this result (circle) and SAPHIR (triangle)~\cite{Lawall:2005np} for two different photon energy
				ranges. The photon energy range in our result in the left plot is $1.0 < E_{\gamma} < 1.55$ GeV.}
			\label{fig:comparisonToSAPHIR}
		\end{center}
	\end{figure*}

	We have measured the $\Sigma^{+}$ transverse polarization ($P_{\Sigma^{+}}$) in photoproduction on a hydrogen target in the photon beam energy
	range 1.0-3.5 GeV (which corresponds to $\sqrt{s} \approx$ 1.66-2.73 GeV). The $\Sigma^{+}$ is significantly polarized in most 
	of the kinematic region and its magnitude goes up to 95\%. 
	Fig.~\ref{fig:comparisonToSAPHIR} shows the comparison of our result with SAPHIR~\cite{Lawall:2005np} for the corresponding kinematic region. 
	Our results are in good agreement with SAPHIR but with better precision.

\begin{figure*}[hbt]
		\begin{center}
			\resizebox{18cm}{!}{\includegraphics*{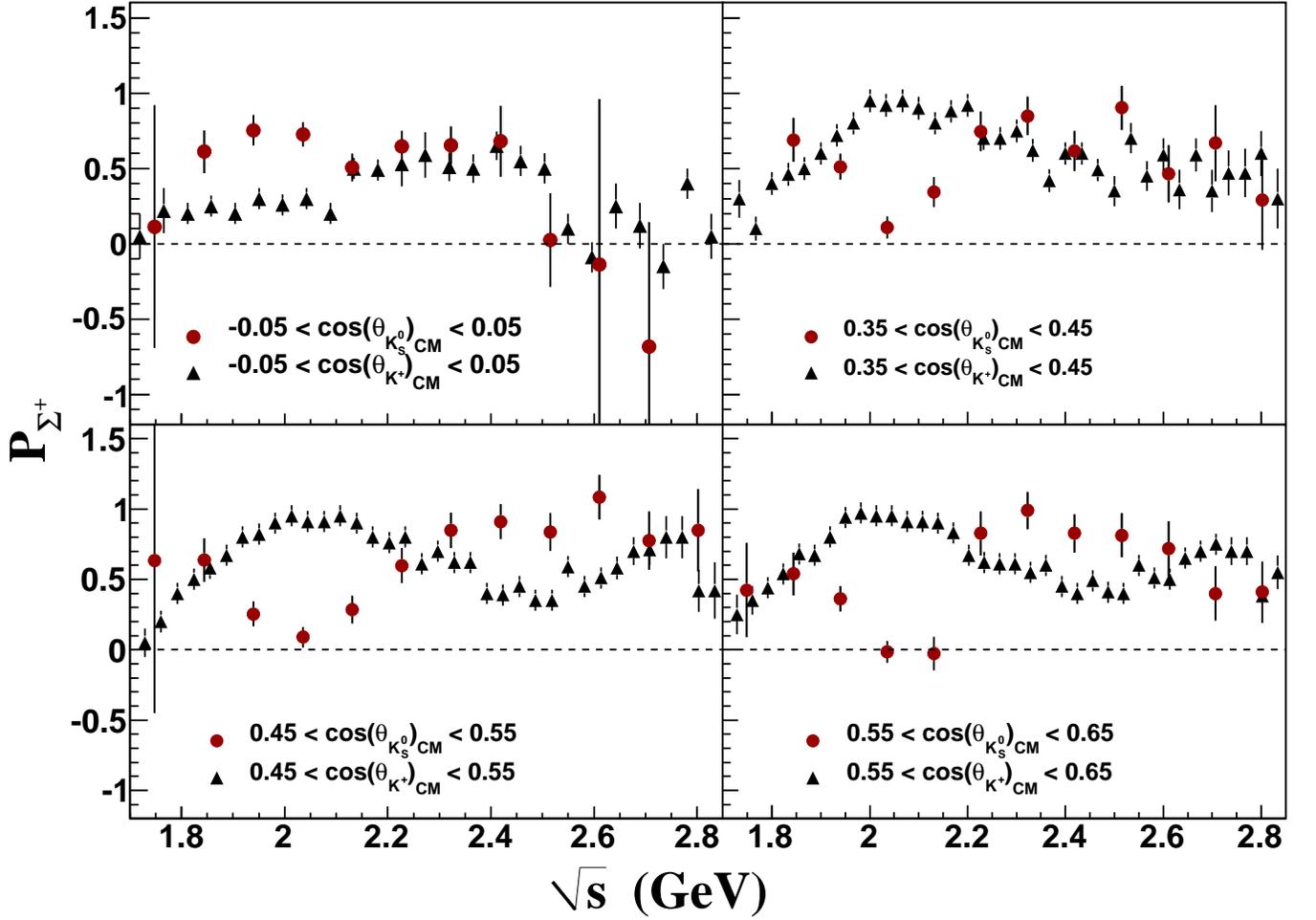}}
			\caption[Comparison of polarization between $\Sigma^{+}$ (this result) and $\Sigma^{0}$] {(Color online) Comparison of $\Sigma^{+}$ 
				polarization from this result (circle, in reaction $\gamma p \ra \Sigma^{+} K^{0}$) with $\Sigma^{0}$ polarization from 
				Ref.~\cite{Dey:2010hh} (triangle, in reaction $\gamma p \ra \Sigma^{0} K^{+}$) for four different angles.} 
			\label{fig:comparisonPlot}
		\end{center}
	\end{figure*}

	SU(6) symmetry and the idea based on a polarization of the $s$ quark~\cite{Heller:1990gc} produced from the sea suggest 
	$P_{\Lambda} \approx -P_{\Sigma^{+}} \approx -P_{\Sigma^{-}} \approx -P_{\Sigma^{0}}$. However, it has been shown in Ref.~\cite{Dey:2010hh} that 
	this symmetry between the $\Lambda (1115)$ and $\Sigma^{0} (1193)$ is broken explicitly in mid and forward angles of the hyperon in the CM frame. 
	Comparison plots of the polarization of the $\Sigma^{+}(1189)$ and the $\Sigma^{0}(1193)$~\cite{Dey:2010hh} are shown in Fig.~\ref{fig:comparisonPlot}.
	For comparison with Ref.~\cite{Dey:2010hh}, we used $\cos(\theta_{\Sigma^{+}})_{\text{CM}} = -\cos(\theta_{K^{+}})_{\text{CM}}$. Also,
	the $\hat{n}_{z} = \frac{\hat{\gamma} \times \hat{K}^{+}}{|\hat{\gamma} \times \hat{K}^{+}|}$ direction is taken as the quantization axis in Ref.~\cite{Dey:2010hh}. 
	Therefore, we have scaled 
	our result by $-1$ (Fig.~\ref{fig:comparisonPlot} only). Because of the low statistics in the forward direction, we compared here data points for 
	backward going $\Sigma^{+}$ only. 
	We can see that the trend of the polarizations with CM energies, $\sqrt{s}$, in both cases is similar except with systematic 
	differences of about 1 at $\sqrt{s} = 2$ GeV. 
	Also, as we can see from Fig.~\ref{fig:polCosDivideAccpCorrBkSubt} that the trend of the polarizations near the resonance regime ($\sqrt{s} \approx 2.0$ GeV) and above
	the resonance regime ($\sqrt{s} \approx 2.5$ GeV) is different. This might indicate that the production mechanisms in these two regimes are different. 
	Recently, several resonances have been shown to exist at around $\sqrt{s} \approx$ 2 GeV~\cite{Anisovich:2011ye,Anisovich:2012ct}. This difference in polarization 
	might be due to the resonance effects of the different contributing s-channel states in these two mass ranges. 

	Because of low statistics, especially at high energy, and for the forward and backward directions, it is difficult to track the variation of the
	polarization with different kinematic variables. For better understanding of the mechanism of polarization in the photoproduction process, 
	and to understand the polarization mechanism at higher energy and at higher transverse momentum ($p_T$), measurements at even higher energies with 
	good statistics are necessary. 
\section{Acknowledgments}
	We would like to acknowledge the outstanding efforts of the staff of the Accelerator and the Physics Divisions at Jefferson Lab that made 
	the experiment possible. This work was supported in part by the Italian Istituto Nazionale di Fisica Nucleare, the French Centre
	National de la Recherche Scientifique and Commissariat \`a l'Energie Atomique, the U.S. Department of Energy and National Science Foundation, 
	the National Research Foundation of Korea, and the United Kingdom's Science and Technology Facilities Council. The Southeastern Universities Research Association (SURA) 
	operates the Thomas Jefferson National Accelerator Facility for the United States Department of Energy under contract DEAC05-84ER40150.
%
%

\end{document}